\documentclass[apj]{emulateapj}
\usepackage{graphicx}
\usepackage{epstopdf}
\usepackage{amsmath}
\usepackage{amssymb}
\shorttitle{Neutron star accretion flow}
\shortauthors{Bu, Qiao, \& Yang}
\begin{document}
\title{Hot accretion flow around neutron stars}
\author{De-Fu Bu\altaffilmark{1}, Erlin Qiao\altaffilmark{2,3} Xiao-Hong Yang\altaffilmark{4}}

\altaffiltext{1}{Key Laboratory for Research in Galaxies and Cosmology, Shanghai Astronomical Observatory, Chinese Academy of Sciences, 80 Nandan Road, Shanghai 200030, China; dfbu@shao.ac.cn }
\altaffiltext{2}{Key Laboratory of Space Astronomy and Technology, National Astronomical Observatory, Chinese Academy of Sciences, Beijing 100012, China; qiaoel@nao.cas.cn }
\altaffiltext{3}{School of Astronomy and Space Sciences, University of Chinese Academy of Sciences, 19A Yuquan Road, Beijing 100049 }
\altaffiltext{4}{Department of physics, Chongqing University, Chongqing, 400044}

%\date{Accepted 1988 December 15. Received 1988 December 14; in original form 1988 October 11}

%\pagerange{\pageref{firstpage}--\pageref{lastpage}} \pubyear{2002}

%\maketitle

%\label{firstpage}
%
\begin{abstract}
We perform as the first time hydrodynamic simulations to study the properties of hot accretion flow (HAF) around a neutron star (NS). The energy carried by the HAF will eventually be radiated out at the surface of the NS. The emitted photons can propagate inside the HAF and cool the HAF via Comptonization. We find that the Compton cooling can affect the properties of HAF around a NS significantly. We define the Eddington accretion rate as $\dot M_{\rm Edd}=10L_{\rm Edd}/c^2$, with $L_{\rm Edd}$ and $c$ being the Eddington luminosity and the speed of light, respectively. We define $\dot m$ as the mass accretion rate at the NS surface in unit of $\dot M_{\rm Edd}$. When $\dot m > 10^{-4}$, Compton cooling can effectively cool the HAF and suppress wind. Therefore, the mass accretion rate is almost a constant with radius. The density profile is $\rho \propto r^{-1.4}$. When $\dot m < 10^{-4}$, the Compton cooling effects become weaker, wind becomes stronger, accretion rate is proportional to $r^{0.3-0.5}$. Consequently, the density profile becomes flatter, $\rho \propto r^{-1 \sim -0.8}$. When $\dot m < 10^{-6}$, the Compton cooling effects can be neglected. We find that with a same accretion rate, the temperature of HAF around a NS is significantly lower than that of HAF around a black hole (BH). Also, the Compton $y-$parameter of HAF around a NS is significantly smaller than that of HAF around a BH. This result predicts that HAF around a NS will produce a softer spectrum compared to HAF around a BH, which is consistent with observations.
\end{abstract}


\section{Introduction}

In black hole low mass X-ray binaries (BH-LMXBs) and neutron star low mass X-ray binaries (NS-LMXBs), the accretor accretes gas from a low mass star. There are two main spectral states for both BH-LMXBs and NS-LMXBs. They are high/soft and low/hard spectral states. There is a critical luminosity ($L_{\rm c}$). When luminosity $L > L_{\rm c}$, the sources are in the high/soft state. When $L < L_{\rm c}$, the sources are in the low/hard state. Previous works show that $L_{\rm c}$ is about $1 \sim 4$ percents of $L_{\rm Edd}$ (e.g., Nowak et al. 2002; Maccarone 2003; Maccarone \& Coppi 2003; Rodriguez et al. 2003; Kubota \& Done 2004;  Meyer-Hofmeister et al. 2005; Gladstone et al. 2007; Qiao \& Liu 2009; Zhang et al. 2016; Zhang \& Yu 2018). In the high/soft state, a geometrically thin and optically thick disk operates (Shakura \& Sunyaev 1973). A thin disk emits a multi-color black body spectrum.

In the low/hard state, a geometrically thick HAF operates (e.g., Narayan \& Yi 1994; Yuan \& Narayan 2014). However, even at a same accretion rate, the properties of HAF around a BH may be different from those of HAF around a NS. The reason is as follows. For a NS, there is a hard surface. Therefore, the energy carried by the accretion flow will eventually be radiated out at the surface of the NS (Zampieri et al. 1995). However, for a BH, the energy carried by the accretion flow can go into the event horizon. The different boundary conditions of BH and NS will result in some significant differences between properties of accretion flows around BH and NS. Observations of NS-LMXBs in the luminosity range $0.01\%L_{\rm Edd}<L<1\%L_{\rm Edd}$ do show that there is a thermal soft X-ray component (e.g., Jonker et al. 2004; Armas Padilla et al. 2013a, 2013b; Degenaar et al. 2013a; Campana et al. 2014). Recently, it is shown that the thermal soft X-ray component is produced at the surface of the NS (Degenaar et al. 2013b;  Hernandez Santisteban et al. 2018; van den Eijnden et al. 2018). In a similar luminosity range, observations of BH-LMXBs do not show such a soft X-ray component (Degenaar et al. 2013a; Bahramian et al. 2014). Observations show that at a same accretion rate in the quiescent state, a NS system is always brighter than a BH system (e.g., Menou et al. 1999; Lasota 2000; Garcia et al. 2001; Hameury et al. 2003; McClintock et al. 2004).
The thermal soft photons from the surface of a NS can propagate outwards and cool the HAF via inverse Compton scattering. Therefore, one would expect that with a same accretion rate, the gas temperature of HAF around a NS should be lower than that of HAF around a BH (Sunyaev \& Titarchuk 1989; Sunyaev et al. 1991; Narayan \& McClintock 2008). Burke et al. (2017) analyzed the X-ray spectrum of a sample composed of 12 BH-LMXBs and NS-LMXBs. It is found that the electron temperature of HAF around a NS is indeed lower than that of a HAF around a BH.

Properties of HAF around a BH have been studied intensively by simulation works (e.g., Stone et al. 1999; Igumenshchev \& Abramowicz 2000; Hawley et al. 2001; Machida et al. 2001;  De Velliers et al. 2003; Pen et al. 2003; Beckwith et al. 2008; Pang et al. 2011; Tchekhovskoy et al. 2011; Narayan et al. 2012; Yuan et al. 2012, 2015; Bu et al. 2016a, 2016b; Bu \& Gan 2018). One of the most important findings of HAF around a BH is that strong wind is present. Due to the presence of strong wind, the mass accretion rate decreases inwards (e.g., Yuan et al. 2012, 2015; Li et al. 2013; Narayan et al. 2012; Bu et al. 2016a, 2016b). Consequently, the radial density profile becomes flatter with $\rho \propto r^{-p}$, with $p<1$. If no wind is present, we will have $\rho \propto r^{-1.5}$. Recently, observations of some low-luminosity active galactic nuclei and hard state of BH-LMXBs show indirect evidences of the presence of wind in HAF around a BH (e.g., Crenshaw \& Kramemer 2012; Wang et al. 2013; Cheung et al. 2016; Homan et al. 2016; Park et al. 2018; Ma et al. 2018).

However, there are no simulations of HAF around a NS. As mentioned above, the soft photons from the surface of a NS can cool the accretion flow via inverse Compton scattering. Therefore, we can expect that the properties of HAF around a NS should be different from those of HAF around a BH. For example, the mass accretion rate (or density) profile of HAF around a NS may be different from that of HAF around a BH. Different mass accretion rate profile will give different emergent spectrum. This is crucial to explain observations of low/hard state of NS-LMXBs. It is very necessary to study the accretion rate profile of HAF around a NS by numerical simulations. Recently, Qiao \& Liu (2018) compared the properties of HAF around NSs and BHs based on one-dimensional self-similar solution. They find that the dynamical properties and emitted spectrums in BH-LMXBs and NS-LMXBs are quite different. The reason for the difference is that for the NS accretion, there is Compton cooling by thermal photons from the surface of the NS. In Qiao \& Liu (2018), they assume that the accretion rate of HAF around both BHs and NSs is a constant with radius.  This assumption should be checked by numerical simulations.

In this paper, we perform simulations of HAF around a NS. In order to compare the results to HAF around a BH, we also perform simulations of HAF around a BH. We study whether wind can be generated in HAF around a NS. Correspondingly, what is the radial profile of mass accretion rate of HAF around a NS? This important property will determine the emergent spectrum of the accreting system.

This paper is organized as follows. In section 2, we introduce the numerical method. In section 3, we present our results; Section 4 is devoted to the discussion. We summarize our results in Section 5.

\section{Numerical method }
We perform two-dimensional simulations of HAF around both NS and BH. We set the black hole mass $M_{\rm BH}=10M_{\odot}$, with $M_{\odot}$ being the solar mass. We set the neutron star mass $M_{\rm NS}=1.4 M_{\odot}$.
We use the ZEUS-MP code (Hayes et al. 2006) and adopt the spherical coordinates ($r,\theta,\phi$) to solve the equations. We list the equations as below,

\begin{equation}
 \frac{d\rho}{dt} + \rho \nabla \cdot {\bf v} = 0,
\end{equation}
\begin{equation}
 \rho \frac{d{\bf v}}{dt} = -\nabla p - \rho \nabla \Phi + \nabla \cdot {\bf T}
\end{equation}
\begin{equation}
 \rho \frac{d(e/\rho)}{dt} = -p\nabla \cdot {\bf v} + {\bf T}^2/\mu + Sc
\end{equation}
$\rho$, $\bf v$ and $e$ are density, velocity and internal energy, respectively. We adopt ideal gas equation $p=(\gamma-1)e$ and set $\gamma=5/3$. The last term in Equation (3) is the inverse Compton cooling of the accretion flow by soft photons from the surface of a NS.

The second term on the right hand side of Equation (2) is gravity. We set the NS gravitational potential $\Phi=GM_{\rm NS}/r$, with $G$ being the gravitational constant. In order to do comparison, for the black hole, we also use Newtonian potential $\Phi=GM_{\rm BH}/r$. We note that for a black hole, pseudo-Newtonian potential is more appropriate, especially inside 10 Schwardzschild radius ($r_s$). However, for the region $r>10r_s$, the pseudo-Newtonian potential is almost same as Newtonian potential.

In Equations (2) and (3), $\bf T$ is the viscous stress tensor. Magnetohydrodynamic simulations (e.g., Stone \& Pringle 2001) find that the azimuthal components of $\bf T$ dominate other components. Therefore, following Stone et al. (1999), we assume that the viscous tensor only has azimuthal components:
\begin{equation}
T_{r\phi} = \mu r \frac{\partial}{\partial r} \left( \frac{v_\phi}{r} \right)
\end{equation}
\begin{equation}
T_{\theta\phi} = \frac{\mu \sin \theta}{r} \frac{\partial}{\partial \theta} \left( \frac{v_\phi}{\sin \theta} \right)
\end{equation}
where, $\mu=\rho \nu$. The viscosity coefficient $\nu \propto c_s^2/\Omega_K$, where, $c_s$ and $\Omega_K$ are sound speed and Keplerican velocity, respectively. For hot accretion flow, $c_s^2$ is proportional to gravitational potential energy. Therefore, $\nu \propto \sqrt{GM} r^{1/2}$ ($M$ is BH or NS mass, Stone et al. 1999). In this paper, following Stone et al. 1999, we assume that $\nu=\alpha \sqrt{GM} r^{1/2}$, which is the usual ``$\alpha$" description and set $\alpha=0.02$.

\begin{table*} \caption{Simulation parameters and results }
\setlength{\tabcolsep}{4mm}{
\begin{tabular}{cccccc}
\hline \hline
 Models & Central object & $\rho_{\rm max}$  & $f_{th}$  & Compton temperature $T_*$ &  Mass inflow rate at  \\

  &   &   ($10^{-8}\text{g cm}^{-3}$) & &  &  inner boundary ($\dot M_{\rm in} (r_{\rm in})/\dot M_{\rm Edd}$)       \\
(1) & (2)             & (3)                         &  (4)      &     (5)  & (6)           \\

\hline\noalign{\smallskip}
BH1   & Black hole & 2.4 & 0 & & $1.8\times 10^{-3}$    \\
NS1   & Neutron star & 1  &  1 & $4.5\times 10^6 $ K & $1.8\times 10^{-3}$   \\
NS2   & Neutron star & 0.1  &  1 & $2.4\times 10^6 $ K & $1.2\times 10^{-4}$   \\
NS3   & Neutron star & 0.02  &  1 & $1.2\times 10^6 $ K & $7 \times 10^{-6}$   \\
NS4   & Neutron star & 0.01  &  1 & $8.1\times 10^5 $ K & $1.3\times 10^{-6}$   \\
NS5   & Neutron star & 1  &  0.25 &  $3.1\times 10^6 $ K & $1.5\times 10^{-3}$   \\
NS6   & Neutron star & 1  &  0.05 &  $1.9\times 10^6 $ K & $8\times 10^{-4}$   \\
NS7   & Neutron star & 1  &  0.01 &  $8.2\times 10^5 $ K & $1.4\times 10^{-4}$   \\

\hline\noalign{\smallskip}
\end{tabular}}

Note: Col. 1: model names. Col. 3: The maximum density at the initial torus center in unit of $10^{-8}\text{g cm}^{-3}$. Cols 4: The ratio of emitted energy per second from NS surface to the total energy transferred to NS surface by HAF per second (see Equations (6) and (7)). Col.5: Compton temperature of photons emitted at NS surface (Equation (7)).  Col. 6: time-averaged mass accretion rate of the central object (in unit of Eddington rate).

\end{table*}

For NSs, a fraction of the energy transferred to the surface of the NSs will be thermalized as blackbody emission. The energy transferred to the surface of a NS per second is,
\begin{equation}
L_*=2 \pi R_*^2 \int_0^\pi \rho_{*} \min [v_{r*},0] \left[ \frac{1}{2}v^2_{*}+e_*/\rho_{*} \right] \sin\theta d\theta
\label{luminosity}
\end{equation}
$\rho_{*}$, $v_{r*}$, $v_{*}$ and $e_{*}$ are density, radial velocity, total velocity, and internal energy at the surface of the NS ($R_*$), respectively. In our simulations, we set the inner boundary to be $10^6$ cm. Therefore, we assume that the physical variables at the inner boundary are equal to those at the surface of the NS. We assume that a fraction ($f_{th}$) of $L_*$ will be thermalized as blackbody emission. As Qiao \& Liu (2018), we assume that the radiation from the NS surface is isotropic, and the effective temperature of the radiation $T_*$ is
\begin{equation}
T_*=\left( \frac{f_{th} L_*}{4\pi R_*^2 \sigma} \right)^{1/4}
\end{equation}
where $\sigma$ is the Stefan-Boltzmann constant. The soft photons emitted from NS surface will propagate outward and cool the accretion flow via Comptonization. The Compton cooling rate in Equation (3) is (Sazonov et al. 2005):
\begin{equation}
Sc=4.1\times10^{-35}n^2(T_*-T)\xi
\label{comptoncooling}
\end{equation}
$n=\rho/(\varrho m_p)$ is the number density of gas, with $\varrho$ and $m_p$ being the mean molecular weight and the proton mass, respectively. We set $\varrho=0.5$. $T$ is the temperature of the accreting gas.  $\xi=f_{th}L_* e^{-\tau'_{es} (r)}/nr^2$ is the ionization parameter, where $\tau'_{es}(r) =\int_0^r \rho \kappa_X dr$ is the X-ray scattering optical depth in radial direction. We set $\kappa_X=0.4 {\rm cm^2 g^{-1}}$.

\subsection{Initial and boundary conditions}

Our computational domain in radial direction is $2.38r_s \leq r \leq 2380 r_s$. For the black hole HAF simulations, $r_s=2GM_{\rm BH}/c^2$; for the NS HAF simulations $r_s=2GM_{\rm NS}/c^2$.  In $\theta$ direction, we have $0 \leq \theta \leq \pi$. Our resolution is $192 \times 88$. In $r$ direction, the grids are logarithmically spaced. In $\theta$ direction, grids are uniformly spaced.

Initially, we put a rotating torus with constant specific angular momentum in our computational domain. The pressure and density in the torus are related through a polytropic equation of state $p=A\rho^\gamma$. The structure of the torus is given by (Papaloizou \& Pringle 1984):
\begin{equation}
\frac{p}{\rho}=\frac{GM}{(n+1)R_0}\left[ \frac{R_0}{r}-\frac{1}{2}\left(\frac{R_0}{r\sin\theta}\right)-\frac{1}{2d} \right]
\end{equation}
In this equation, $M$ is the mass of the central object. In simulations of NS HAF, $M=M_{\rm NS}$. In simulations of BH HAF, $M=M_{\rm BH}$. $R_0$ is the location of torus center.  $R_0$ is located at 470 $r_s$. For the black hole HAF simulations, $R_0=940GM_{\rm BH}/c^2$; for the NS HAF simulations $R_0=940GM_{\rm NS}/c^2$.  $n=(\gamma-1)^{-1}$ is the polytropic index. $d=1.125$ is the distortion parameter of the torus.
The maximum density of torus center is $\rho_{\rm max}$. We assume that the torus is embedded in a non-rotating low-density medium with density $\rho_0=10^{-5}\rho_{\rm max}$. The pressure of the medium is $p=\rho_0/r$.

At the inner and outer radial boundaries, we use outflow boundary conditions. For the outflow boundary conditions, gas is not allowed to flow into the computational domain. The physical variables in the ghost zones are set to be equal to those in the fist active zone. At the rotational axis, axis-of-symmetry boundary conditions are applied.

\begin{figure*}
\begin{center}
\includegraphics[scale=0.5]{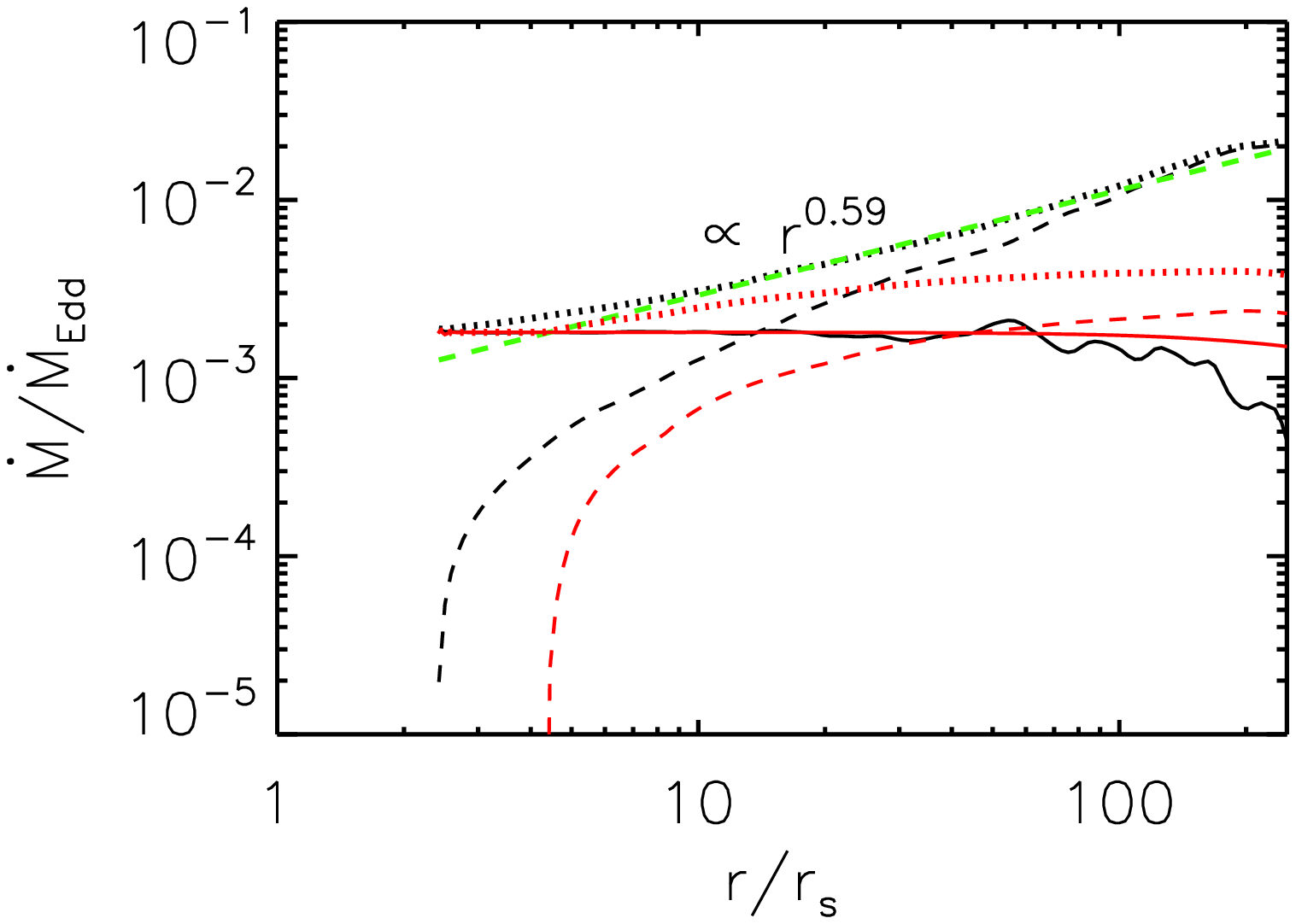}\hspace*{0.7cm}
\includegraphics[scale=0.5]{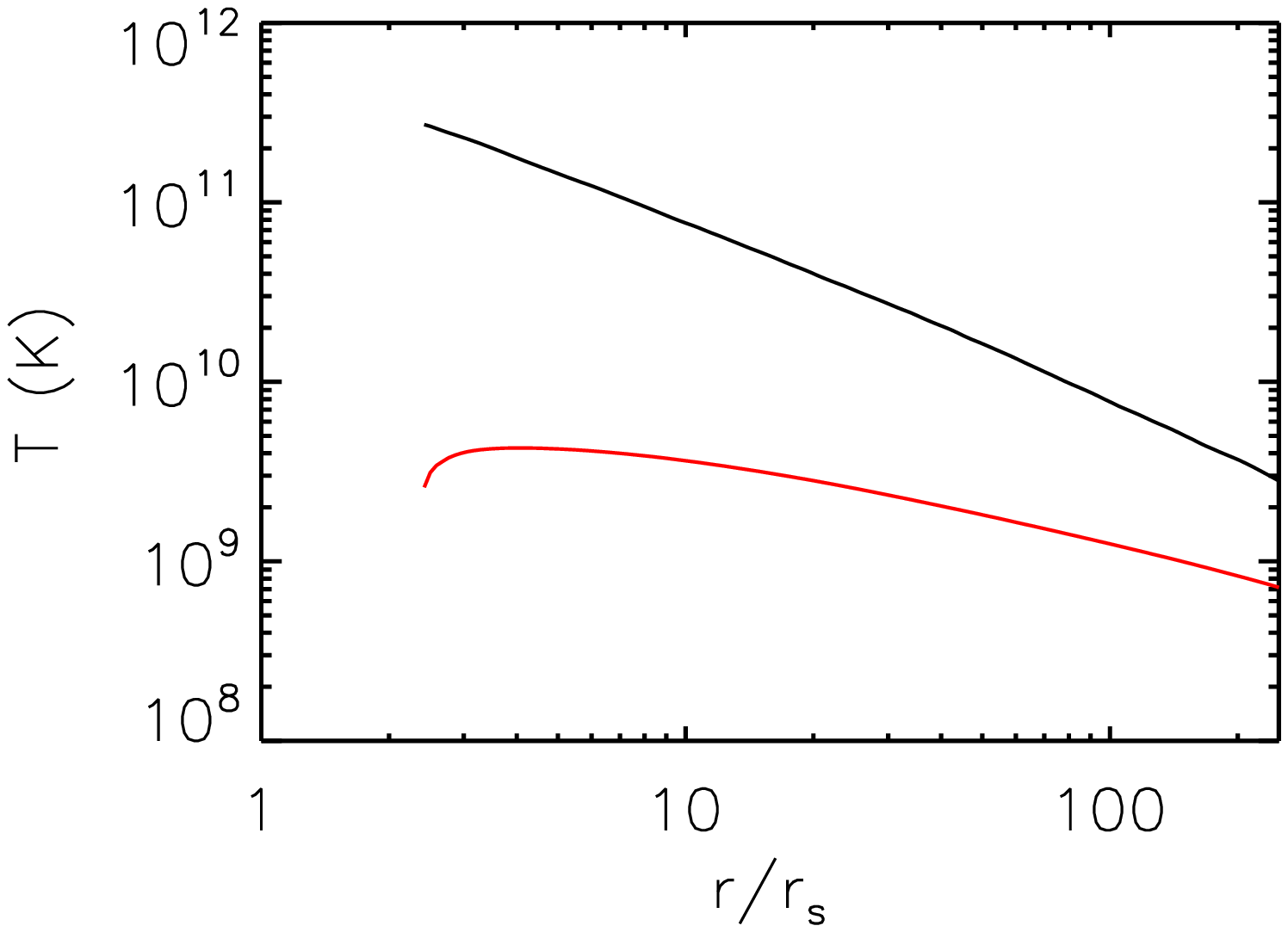}\hspace*{0.0cm} \\
\includegraphics[scale=0.5]{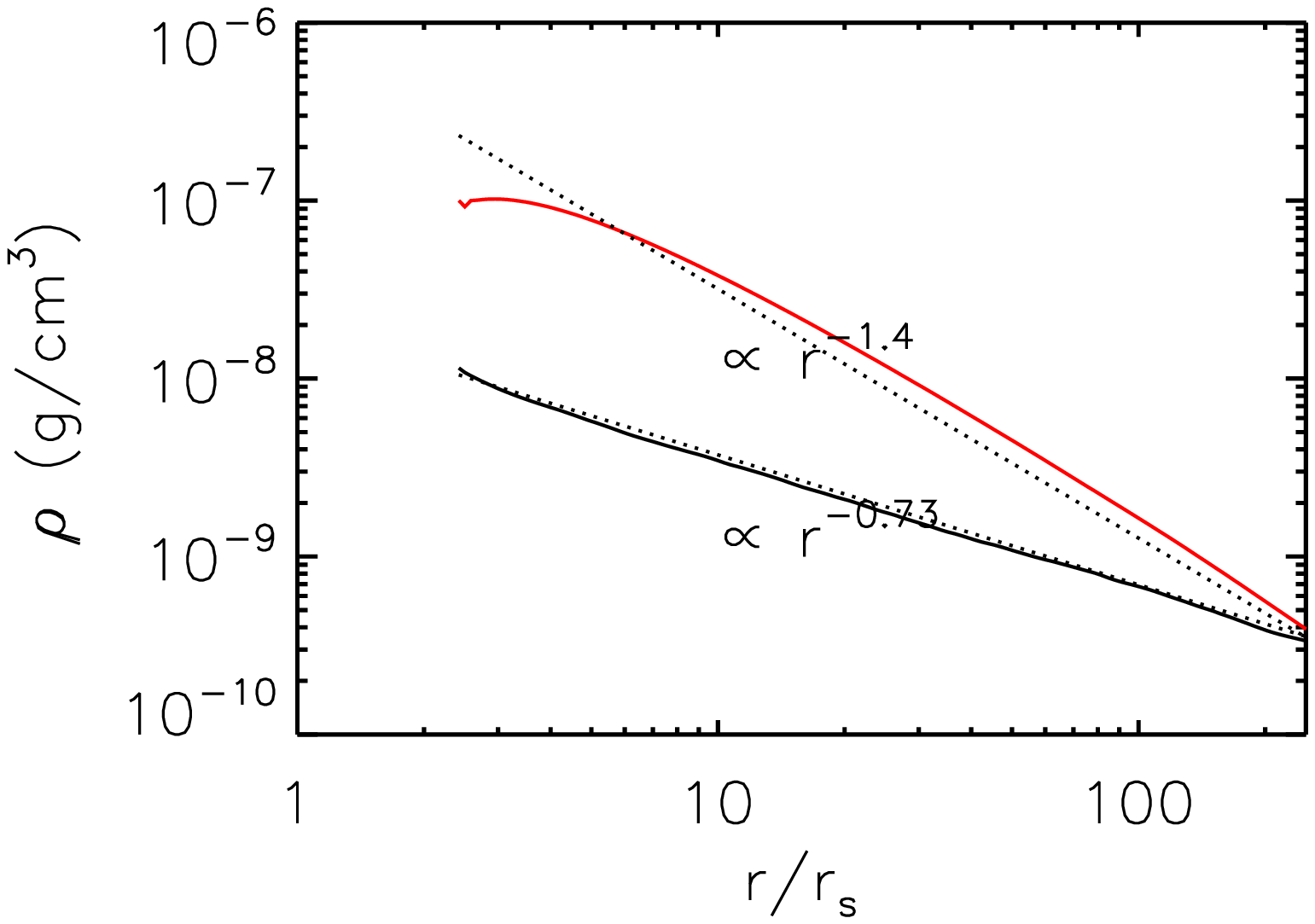}\hspace*{0.7cm}
\includegraphics[scale=0.5]{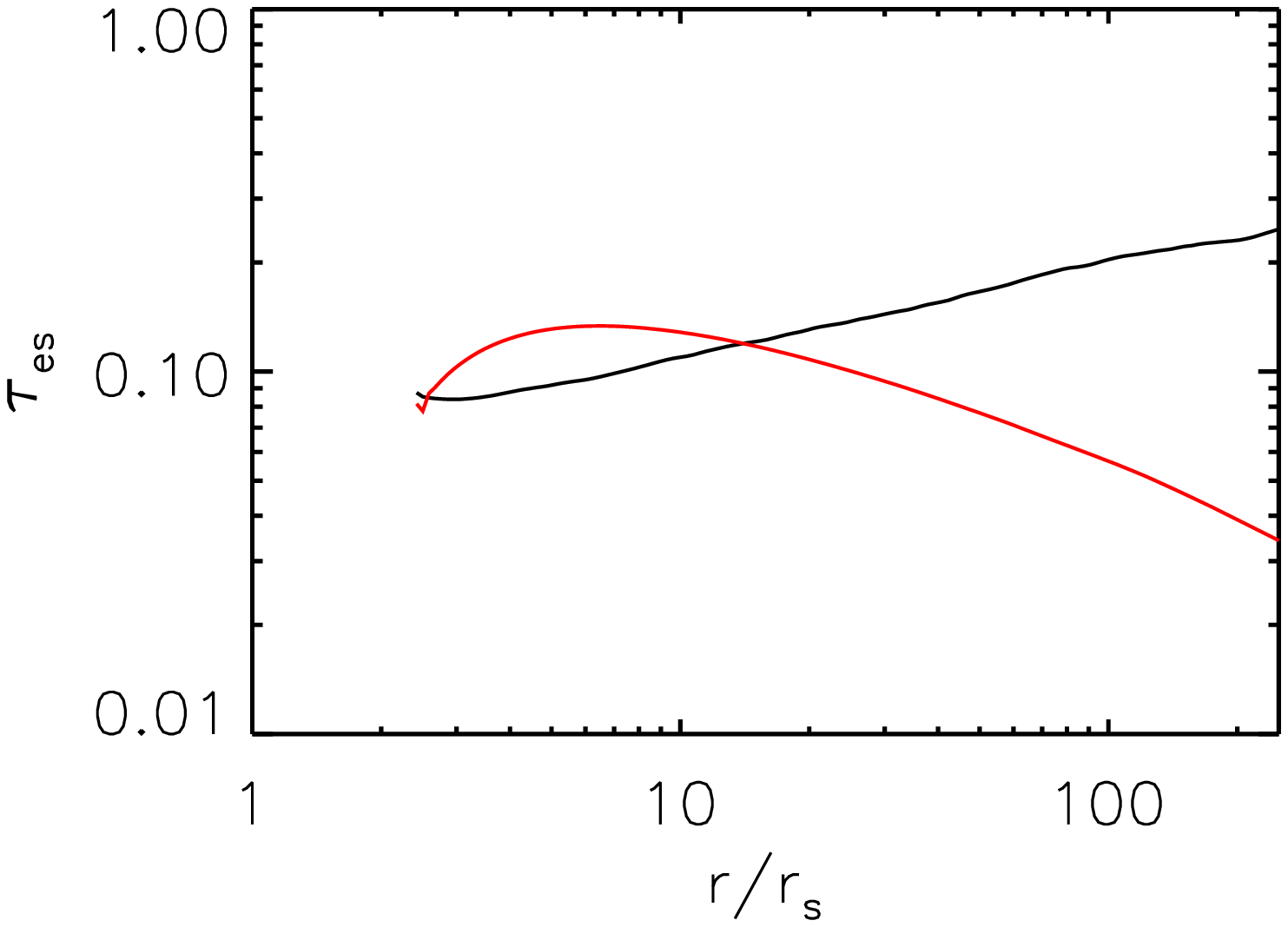}\hspace*{0.0cm} \\
\includegraphics[scale=0.5]{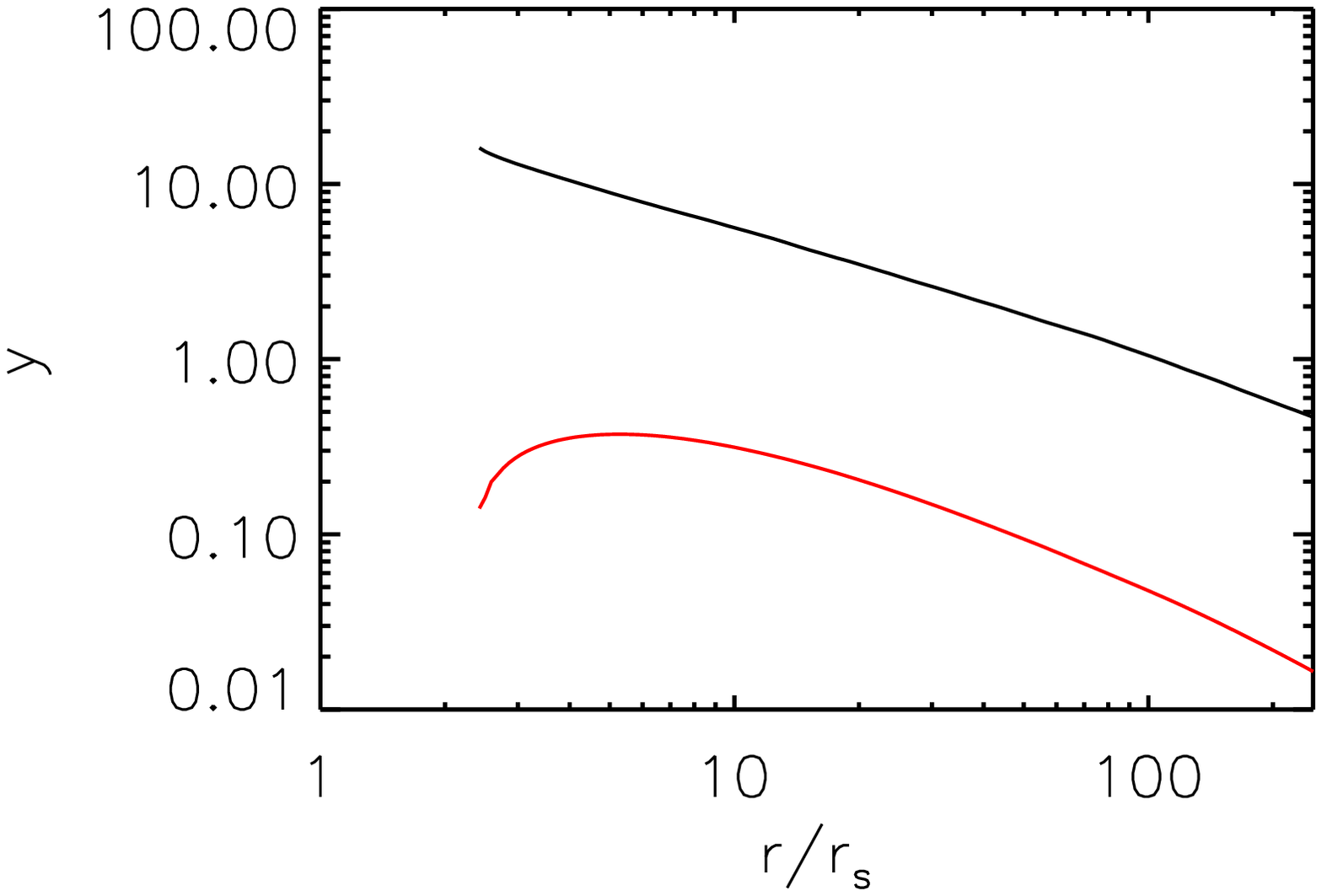}\hspace*{0.7cm}

\hspace*{0.5cm} \caption{Time-averaged (from $t=3$ to 5 orbital time at initial torus center) variables for models BH1 and NS1. In all the panels, black lines are for model BH1, red lines are for model NS1. Top left panel: radial profiles of mass inflow (dotted lines, Equation (\ref{inflowrate})), outflow (dashed lines, Equation (\ref{outflowrate})) and net (solid lines, Equation (\ref{netrate})) rates. The green dashed line is the power law function fit to the mass inflow rate of model BH1. Top right panel: mass weighted gas temperature as a function of radius. Middle left panel: radial profile of volume weighted density. The two dotted lines are power law function fit to the density profiles. Middle right panel: Compton scattering optical depth as a function of radius (Equation (\ref{tao})). Bottom panel: Compton $y-$ parameter as a function of radius. \label{Fig:BHvsNS}}
\end{center}
\end{figure*}

\section{Results}
In this paper, we express time in unit of orbital time at initial torus center.
Following Stone et al. (1999), we calculate the mass inflow, outflow and net rates as follows,
(1) inflow rate (or accretion rate)
\begin{equation}
\dot {M}_{\rm in} (r)=-2\pi r^2 \int_{\rm 0}^{\rm \pi}
\rho \min (v_r, 0) \sin\theta d\theta
\label{inflowrate}
\end{equation}
(2) outflow rate
\begin{equation}
\dot {M}_{\rm out} (r)=2\pi r^2 \int_{\rm 0}^{\rm \pi}
\rho \max (v_r, 0) \sin\theta d\theta
\label{outflowrate}
\end{equation}
and (3) net rate
\begin{equation}
\dot {M}_{\rm net} (r)=\dot {M}_{\rm in} (r) - \dot {M}_{\rm out} (r)
\label{netrate}
\end{equation}

The model parameters and some results are summarized in Table 1. We define $\dot M (\rm R_*)$ as the mass accretion rate at the BH horizon or NS surface. We set $\dot M (\rm R_*)$ equaling to the mass inflow rate (calculated in Equation (\ref{inflowrate})) at the inner radial boundary, $\dot M(\rm R_*)= \dot M_{\rm in} (\rm r_{in})$. $\dot M (\rm R_*)$ can be expressed in unit of Eddington rate as $\dot m = \dot M (\rm R_*)/\dot M_{\rm Edd}$.

\begin{figure*}
\begin{center}
\includegraphics[scale=0.5]{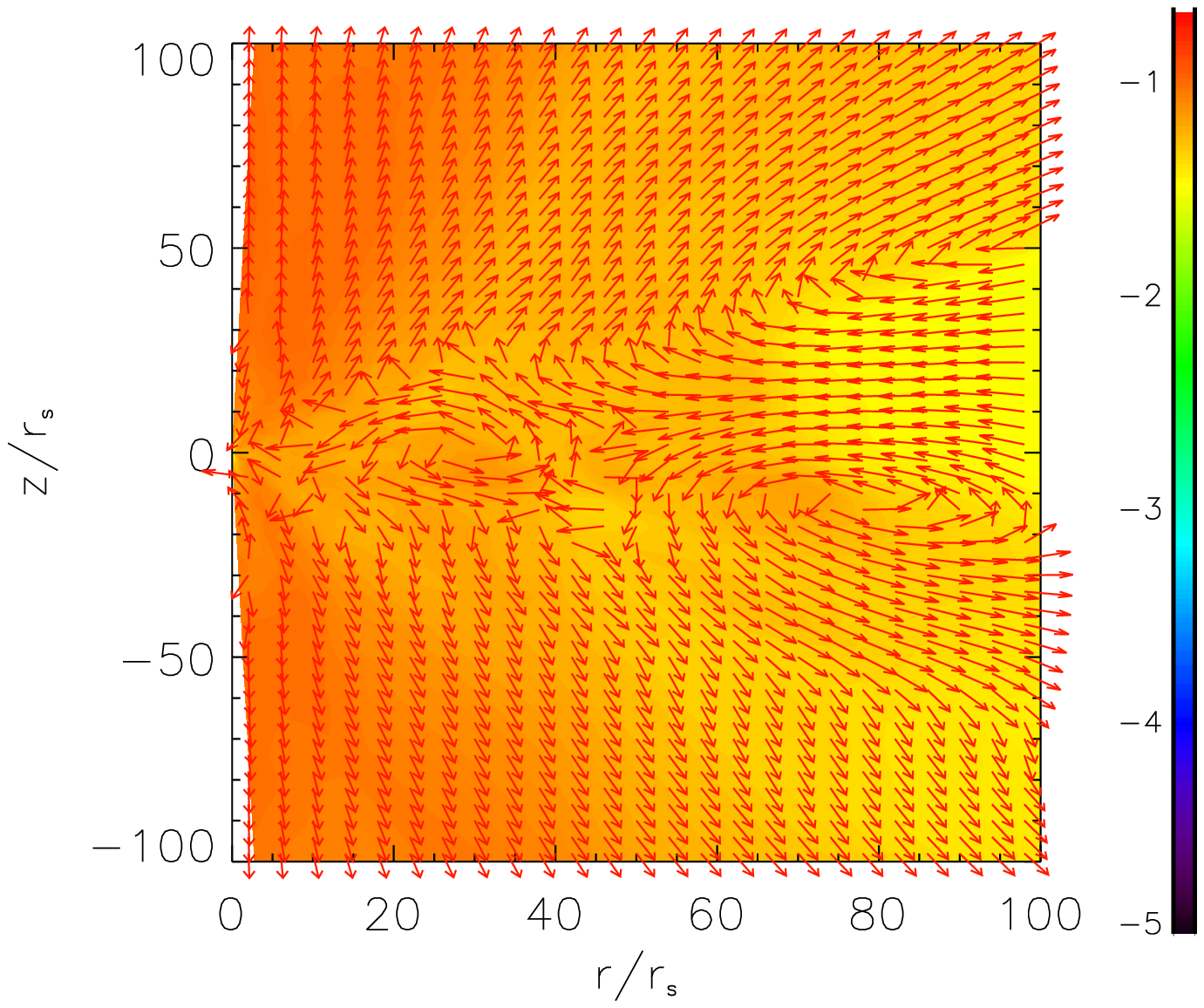}\hspace*{0.0cm}
\includegraphics[scale=0.5]{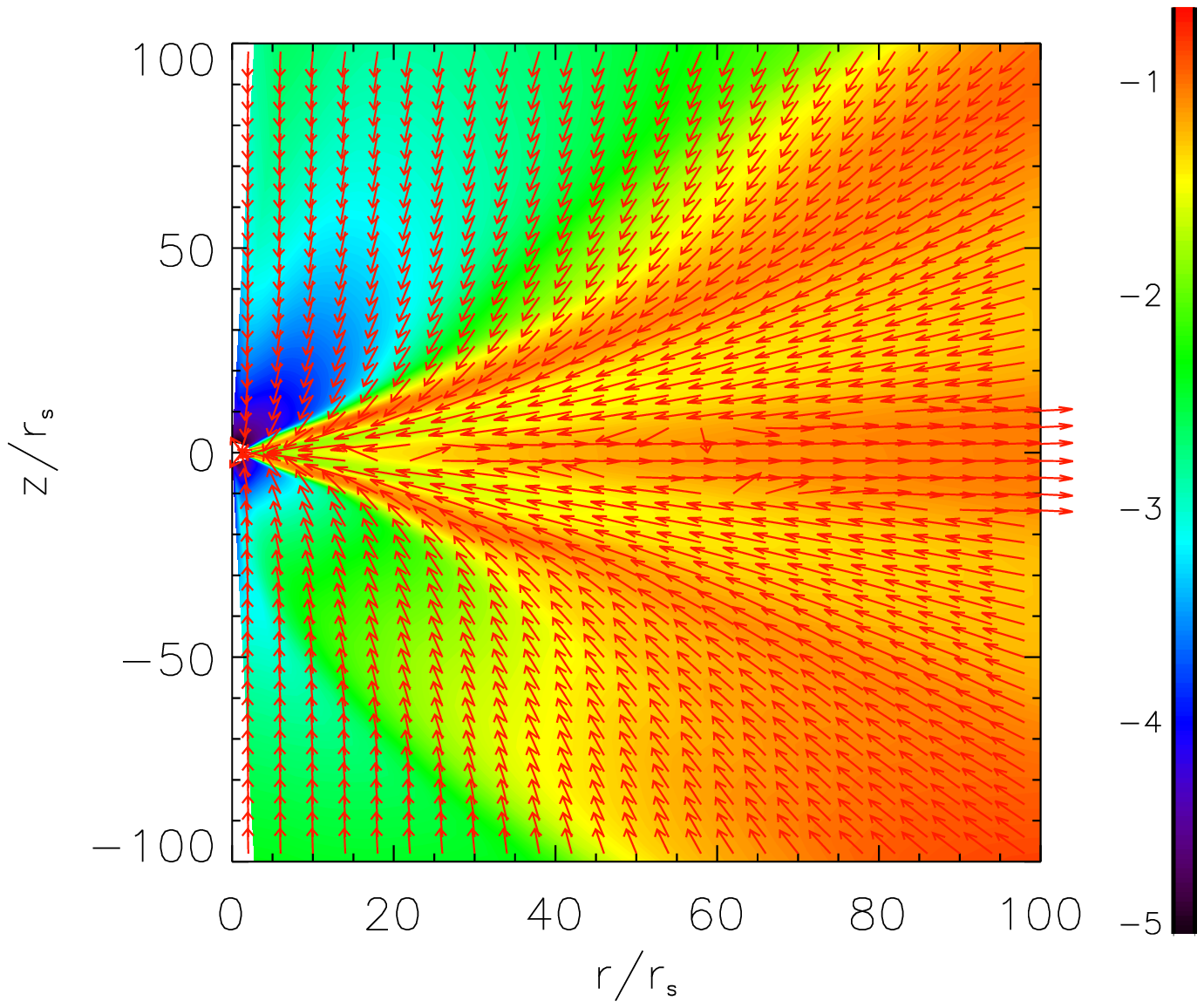}\hspace*{0.7cm}
\hspace*{0.5cm} \caption{Time-averaged (from $t=3$ to 5 orbital time at initial torus center) logarithm temperature in unit of virial temperature (color) and unit velocity vectors for models BH1 (left panel) and NS1 (right panel). \label{Fig:vectorBHvsNS}}
\end{center}
\end{figure*}

\begin{figure}
\begin{center}
\includegraphics[scale=0.5]{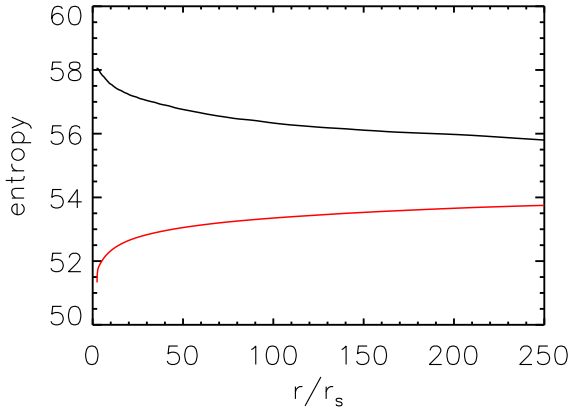}\hspace*{0.0cm}
\hspace*{0.5cm} \caption{Time (from $t=3$ to 5 orbital time at initial torus center) and $\theta$ angle averaged gas entropy as a function of radius for models BH1 (black line) and NS1 (red line). \label{Fig:convection}}
\end{center}
\end{figure}

\begin{figure*}
\begin{center}
\includegraphics[scale=0.5]{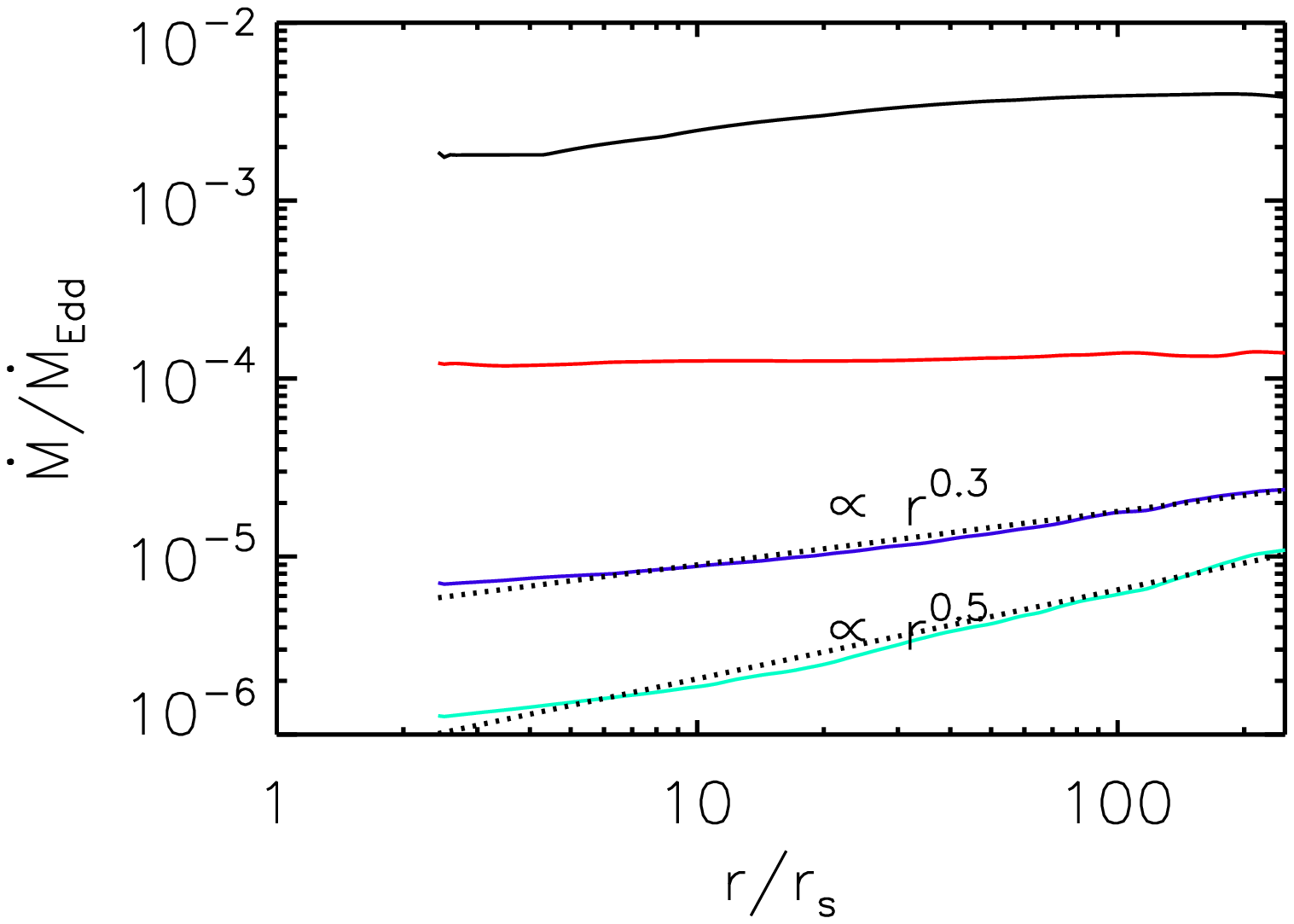}\hspace*{0.7cm}
\includegraphics[scale=0.5]{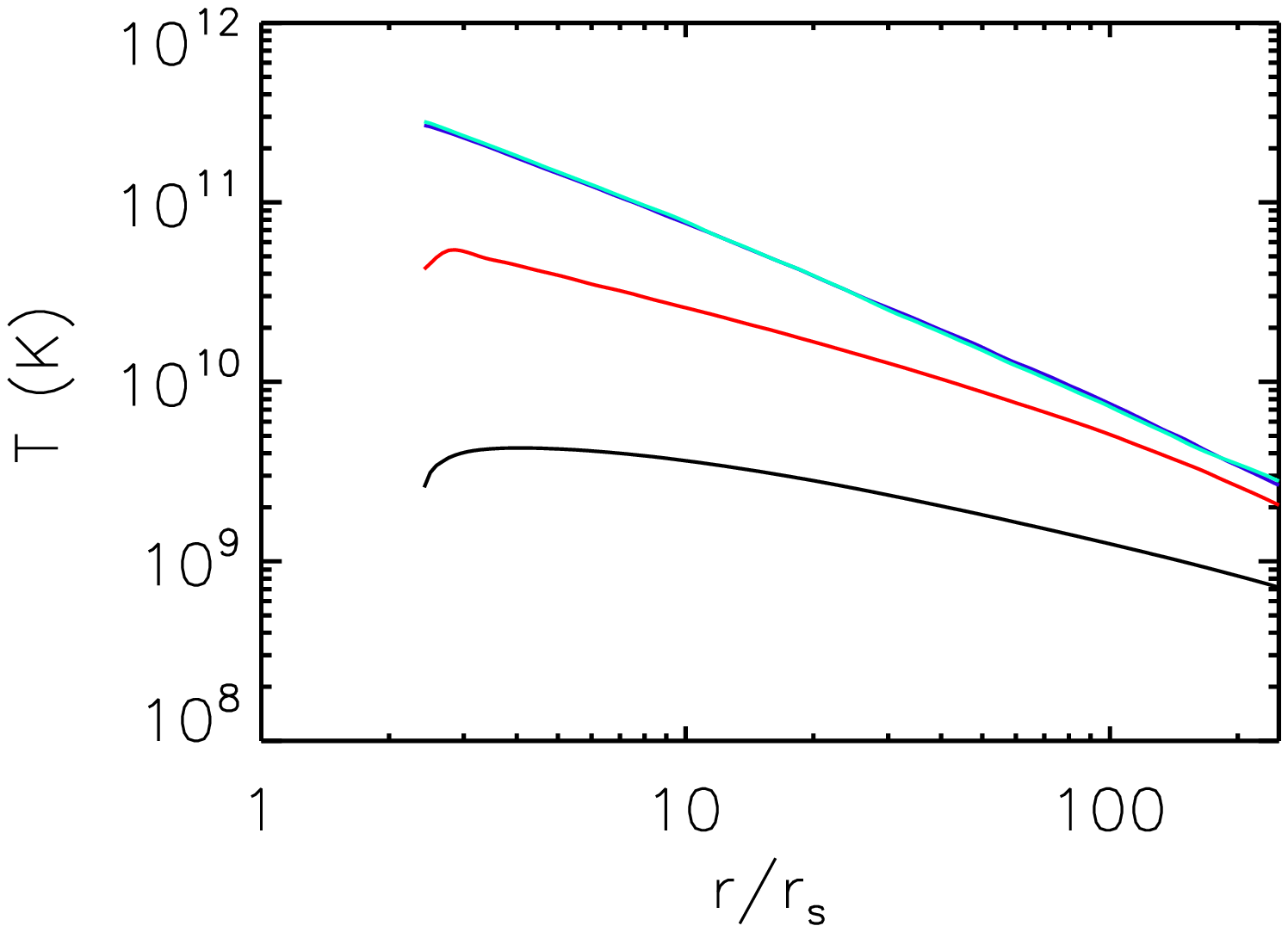}\hspace*{0.0cm} \\
\includegraphics[scale=0.5]{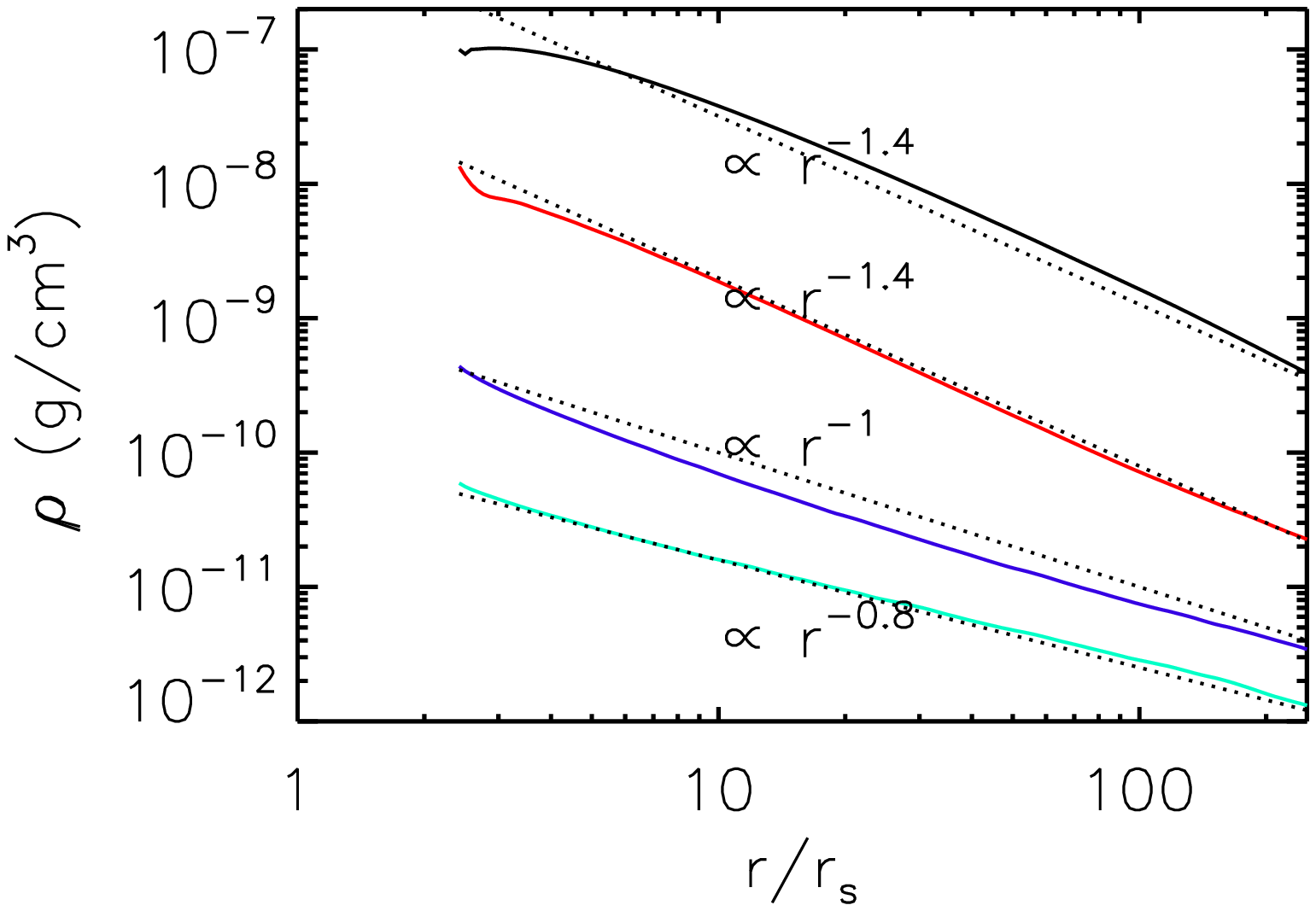}\hspace*{0.7cm}
\includegraphics[scale=0.5]{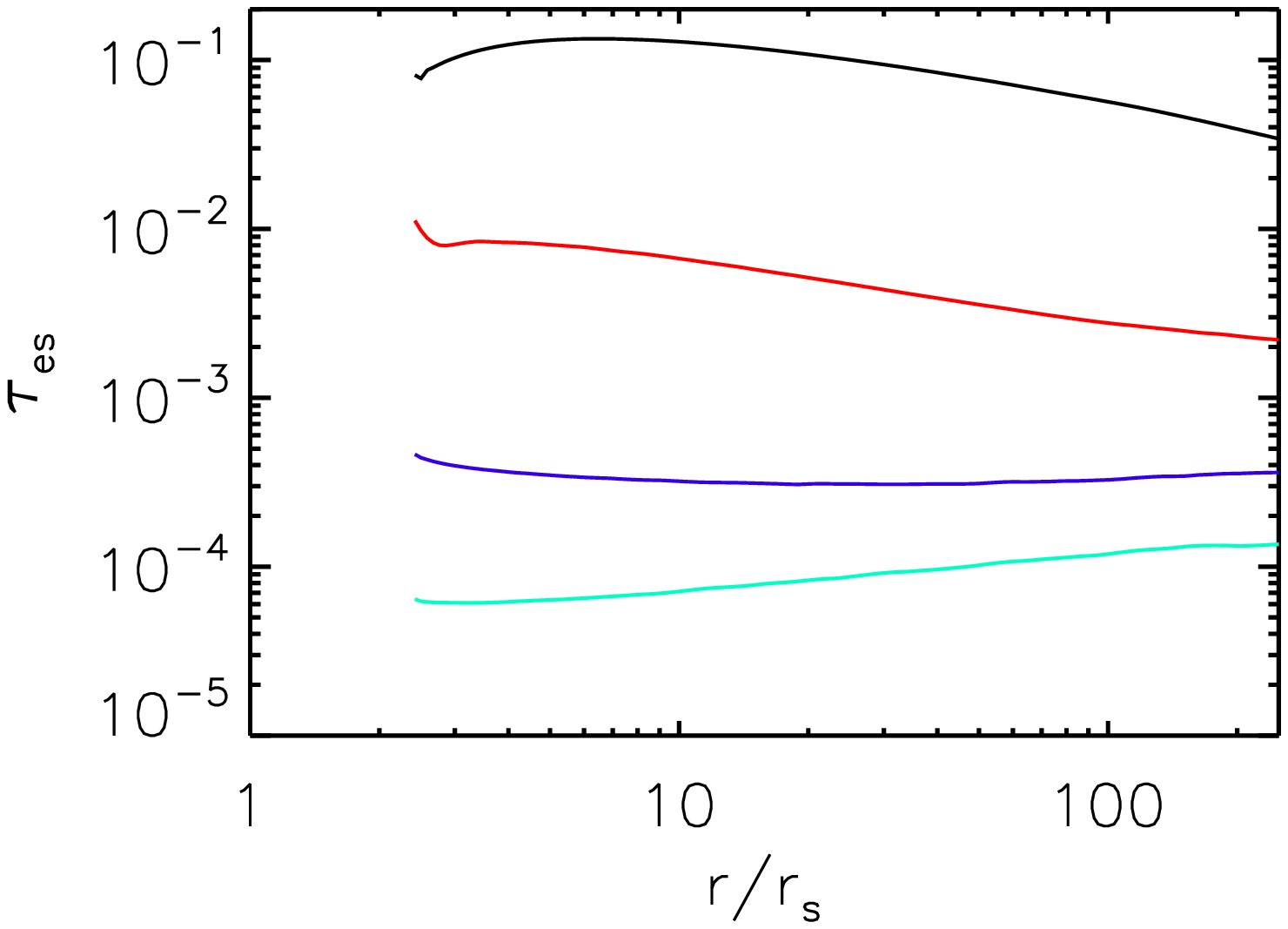}\hspace*{0.0cm} \\
\includegraphics[scale=0.5]{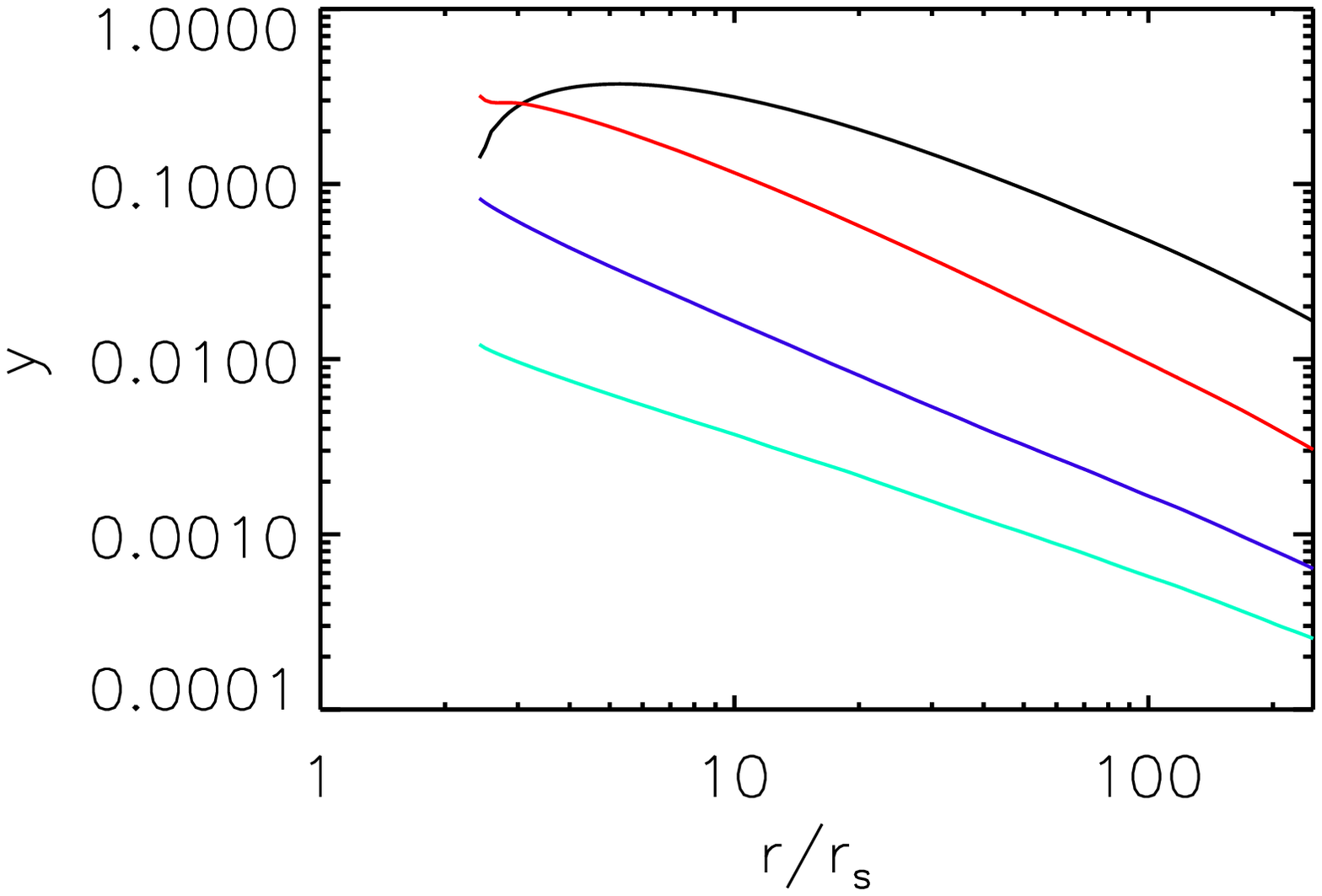}\hspace*{0.7cm}

\hspace*{0cm} \caption{Time-averaged (from $t=3$ to 5 orbital time at initial torus center) variables for models NS1 (black lines), NS2 (red lines), NS3 (blue lines) and NS4 (green lines). Top left panel: radial profiles of mass inflow (see Equation (\ref{inflowrate})) rate. The two dotted lines are power law function fit to the mass inflow rate profiles of models NS3 and NS4. Top right panel: mass weighted gas temperatures as a function of radius. Middle left panel: radial profiles of volume weighted density. The four dotted lines are power law function fit to the density profiles. Middle right panel: Compton scattering optical depths as a function of radius (Equation (\ref{tao})). Bottom panel: Compton $y-$ parameters as a function of radius. \label{Fig:NSmdot}}
\end{center}
\end{figure*}

\subsection{Black hole versus neutron star}
We compare the models BH1 and NS1. The results are shown in Figure \ref{Fig:BHvsNS}.  In this figure, we show time-averaged (from $t=3$ to 5 orbital time at initial torus center) variables as a function of radius for models BH1 and NS1. We find that after 3 orbital times, the net accretion rate is almost a constant with radius (see top-left pane of Figure \ref{Fig:BHvsNS}). Therefore, the results below are given when a quasi-steady state is achieved. In these two models, the mass accretion rate at the central object surface are same. It is $\dot m=1.8\times 10^{-3}$. From the top-left panel, we see that for the BH case, the mass inflow rate decreases inwards (black dotted line). The green dashed line is a power law function fit to the mass inflow rate. The mass inflow rate in the region $10r_s<r<100r_s$ can be described as a power law function of radius $\dot M_{\rm in} \propto r^s$, with $s=0.59$. The result is consistent with that found in previous simulation works studying HAF around a BH (e.g., Stone et al. 1999; Sadowski et al. 2013; Yuan et al. 2012, 2015; Bu et al. 2013), which found that $0.5<s<1$. As explained in previous works (e.g., Stone et al. 1999; Sadowski et al. 2013; Yuan et al. 2012, 2015; Bu et al. 2013), the inward decrease of mass inflow rate is due to the presence of wind (black dashed line in top-left panel of Figure \ref{Fig:BHvsNS}). In Figure \ref{Fig:vectorBHvsNS}, we plot the time-averaged two-dimension distribution of gas temperature in unit of Virial temperature over-plotted with unit velocity vector for models BH1 (left panel) and NS1 (right panel). From the left panel of this figure, we see that winds are present. The gas temperature in the wind region is just slightly lower than Virial temperature.

For model NS1, the mass inflow rate is almost a constant of radius. The value of mass inflow rate at 200$r_s$ is just 2 times that at the NS surface. In the right panel of Figure \ref{Fig:vectorBHvsNS}, we plot the two-dimension structure of this model. It is clear, wind is almost absent. The gas temperature in the regions $0^\circ < \theta < 38^\circ$ and $142^\circ < \theta < 180^\circ$ is significantly lower than that at the same region in model BH1. In the top-right panel of Figure \ref{Fig:BHvsNS}, we plot the mass weighted gas temperatures ($2 \pi \int \rho T r^2 \sin \theta d\theta/2 \pi \int \rho r^2 \sin \theta d\theta $) in models BH1 and NS1 as a function of radius. The gas temperature of BH accretion case is significantly higher than that for the NS case. The Compton cooling by the soft photons from the NS surface significantly decreases gas temperature. The gas temperature in model NS1 is too low. Gas specific energy is too low to generate wind.

From Figure \ref{Fig:vectorBHvsNS}, we see that in model BH1 the convective motions are evident. In this model, the viscous dissipation energy is stored in gas as entropy ($\ln{p/\rho^\gamma}$). The entropy of gas increases inwards (see black line in Figure \ref{Fig:convection} ). Therefore, the accretion flow in model BH1 is convectively unstable (Narayan \& Yi 1994; Yuan \& Bu 2010). For model NS1, due to the strong Compton cooling, the gas entropy decreases inwards (see red line in Figure \ref{Fig:convection} ). Therefore, the accretion flow in model NS1 is convectively stable. There are no convective motions in this model (see vectors in right panel of Figure \ref{Fig:vectorBHvsNS}).

In the middle left panel of Figure \ref{Fig:BHvsNS}, we plot the volume weighted radial profile of density.
\begin{equation}
\rho (r) = \frac{2 \pi r^2 \int_{\rm 0^\circ}^{\rm 180^\circ} \rho \sin \theta d\theta}{2 \pi r^2 \int_{\rm 0^\circ}^{\rm 180^\circ} \sin \theta d\theta}
\label{density}
\end{equation}
In model BH1, strong wind can take away mass, therefore, the density increases slower inwards. The density profile in model BH1 can be described as $\rho \propto r^{-0.73}$. This is also consistent with previous simulations of HAF around a BH (e.g., Stone et al. 1999). In model NS1, wind is very weak (see red dashed line in top left panel of Figure \ref{Fig:BHvsNS}). The radial profile of density is $\rho \propto r^{-1.4}$. If wind is completely absent, we will have $\rho \propto r^{-1.5}$ (e.g., Narayan \& Yi 1994). Therefore, the radial density profile in model NS1 is quite similar as that in models without wind.

In analytical works calculating the spectrum of HAF, one of the most important parameter is the electron scaterring optical depth in vertical direction $\tau_{es}$. In this paper, we define it as follows,
\begin{equation}
\tau_{es}=\int_{\rm 0}^{\rm \pi} \rho \kappa r d\theta
\label{tao}
\end{equation}
In the middle right panel of Figure \ref{Fig:BHvsNS}, we plot the radial profiles of $\tau_{es}$. In the region $2.3r_s < r < 10r_s$ where most of the photons are radiated out by a HAF, the Compton scattering optical depth $\tau_{es}$ in model NS1 is higher. However, outside $10r_s$, $\tau_{es}$ in model BH1 is higher. $\tau_{es}$ is proportional to $\rho$, the density in model NS1 in higher than that in model BH1 at all radii (see middle left panel of Figure \ref{Fig:BHvsNS}). Therefore, one would expect that $\tau_{es}$ in model NS1 should be larger than that in model BH1 at all radii. We note that the density profile is volume weighted (see Equation \ref{density}). There is a geometric factor $\sin \theta$ when we calculate density profile. However, when we calculate the $\tau_{es}$ (Equation (\ref{tao})), the geometric factor does not exist.

The Compton $y-$parameter is another important parameter when one calculates the spectrum of a accretion flow. It is defined as follows,
\begin{equation}
y=\frac{4 k T}{m_e c^2} \max(\tau_{es}, \tau_{es}^2)
\end{equation}
where $k$ and $m_e$ are Boltzmann constant and electron mass, respectively. In this equation, in principal, we should use electron temperature. However, we use one-temperature simulations. Therefore, we use $T$ instead of electron temperature. The gas temperature in model NS1 is significantly lower than that in model BH1 (top right panel of Figure \ref{Fig:BHvsNS}). Also, $\tau_{es}$ is only slightly larger in model NS1 than that in model BH1 in the region $r<10r_s$. Therefore, the Compton $y-$parameter in model BH1 is significantly larger than that in model NS1 at all radii. This is consistent with that found in analytical work (Qiao \& Liu 2018). If accretion rate is $\sim 10^{-3} \dot M_{\rm Edd}$, the results predict that HAF around a BH can produce a harder X-ray spectrum than that of HAF around a NS.

\subsection{HAFs around a NS with different accretion rates}
We carry out models NS1, NS2, NS3 and NS4 (see Table 1) to study the properties of HAFs around a NS with different accretion rates. In these four models, we set $f_{th}=1$. The results are shown in Figure \ref{Fig:NSmdot}. From Equation (\ref{comptoncooling}), we now that Compton cooling rate $Sc \propto n L_*$. Because $L_* \propto n$ (Equation (\ref{luminosity})), so we have $Sc \propto n^2$. The Compton cooling timescale is $e/Sc \propto n^{-1} \propto \dot m^{-1}$. The gas infall timescale is $r/v_r$. For HAF, gas infall velocity is proportional to Keplerian velocity and does not change much with varying accretion rate (e.g., Narayan \& Yi 1994). Therefore, gas infall timescale does not change much with varying accretion rate. However, the Compton cooling timescale is inversely proportional to accretion rate. Therefore, we can expect that Compton cooling effects will become smaller and smaller with decreasing mass accretion rate.

The top left panel of Figure \ref{Fig:NSmdot} shows the radial profiles of mass inflow rate. The two dotted lines in this panel are power law function fit to the mass inflow rate profiles for models NS3 and NS4. With the decrease of $\dot m$, effects Compton cooling become smaller, winds become stronger. For models with $\dot m > 10^{-4}$, wind is very weak, the mass inflow rate is approximately a constant with radius. In model NS3, due to the presence of wind, $\dot m$ at NS surface is $28\%$ of that at 250 $r_s$ and we have $\dot M_{\rm in} \propto r^{0.3}$. In model NS4, this ratio decreases to $12\%$ and we have $\dot M_{\rm in} \propto r^{0.5}$.  From the middle left panel, we see that with the decrease of $\dot m$, the radial density profile becomes flatter. When $\dot m < 10^{-6}$, the effects of Compton cooling can be neglected. In this case, the properties of HAF around a NS are same as those of HAF around a BH. With the decrease of $\dot m$, the gas temperature increases (top right panel of Figure \ref{Fig:NSmdot}). In the middle right panel, we plot the Compton scattering optical depth ($\tau_{es}$) as a function of radius. $\tau_{es}$ decreases with decreasing $\dot m$, as predicted by the formula of $\tau_{es}$ (Equation (\ref{tao})). The bottom panel shows the radial profiles of Compton $y-$parameter. As found by Qiao \& Liu (2018), it decreases with decreasing $\dot m$. The X-ray spectrum will become softer with decreasing of $\dot m$.

\begin{figure*}
\begin{center}
\includegraphics[scale=0.5]{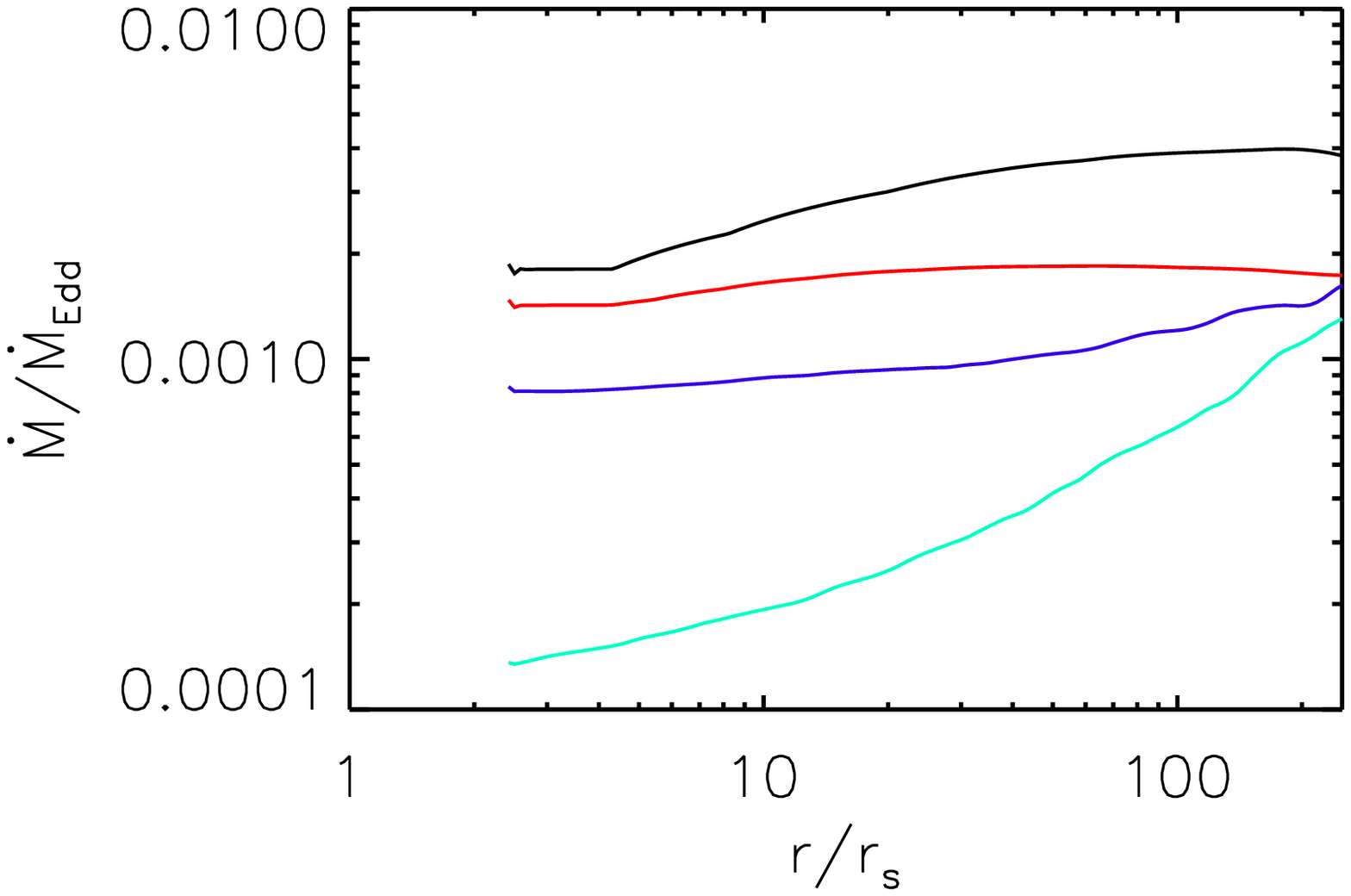}\hspace*{0.7cm}
\includegraphics[scale=0.5]{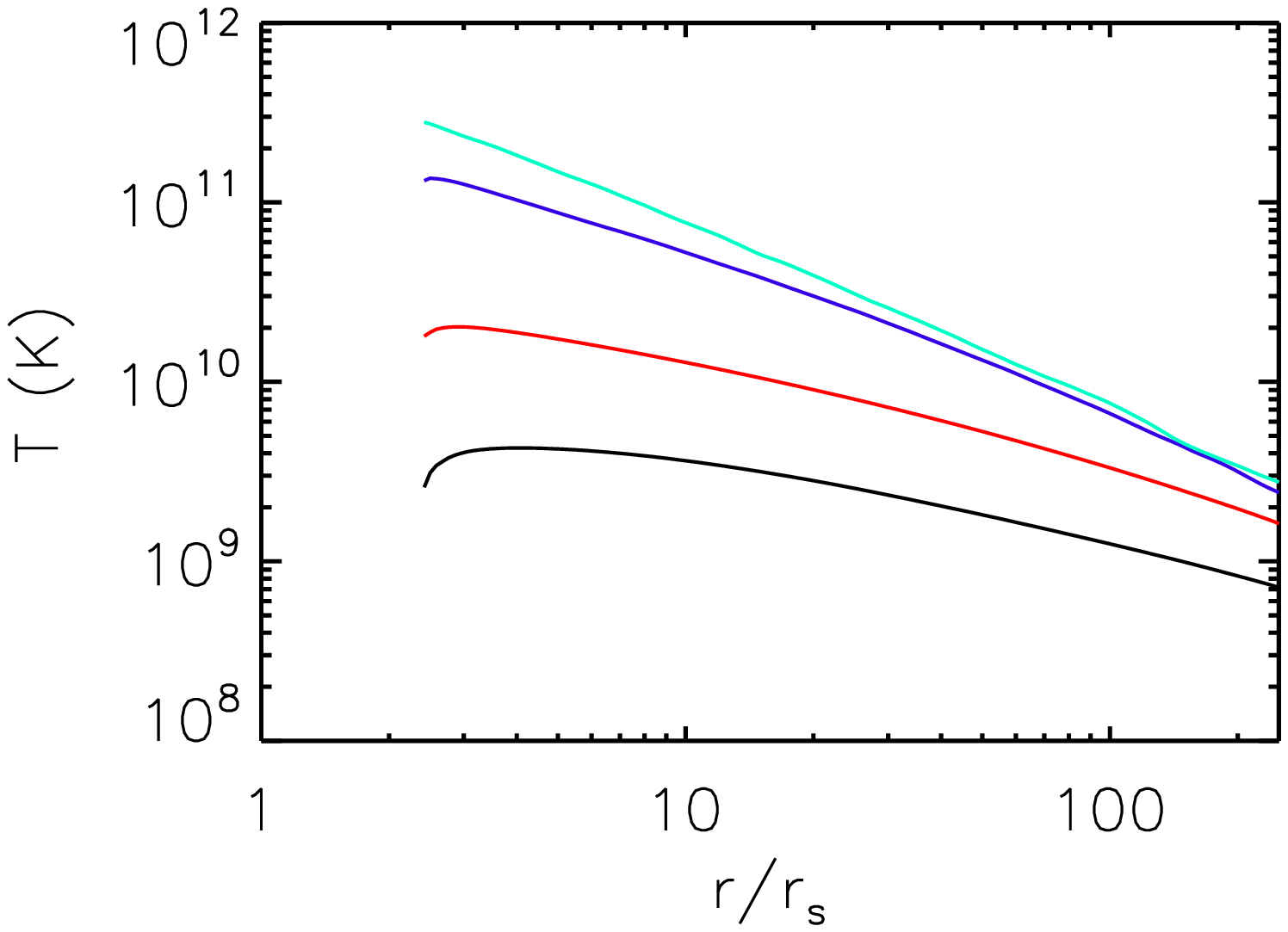}\hspace*{0.0cm} \\
\includegraphics[scale=0.5]{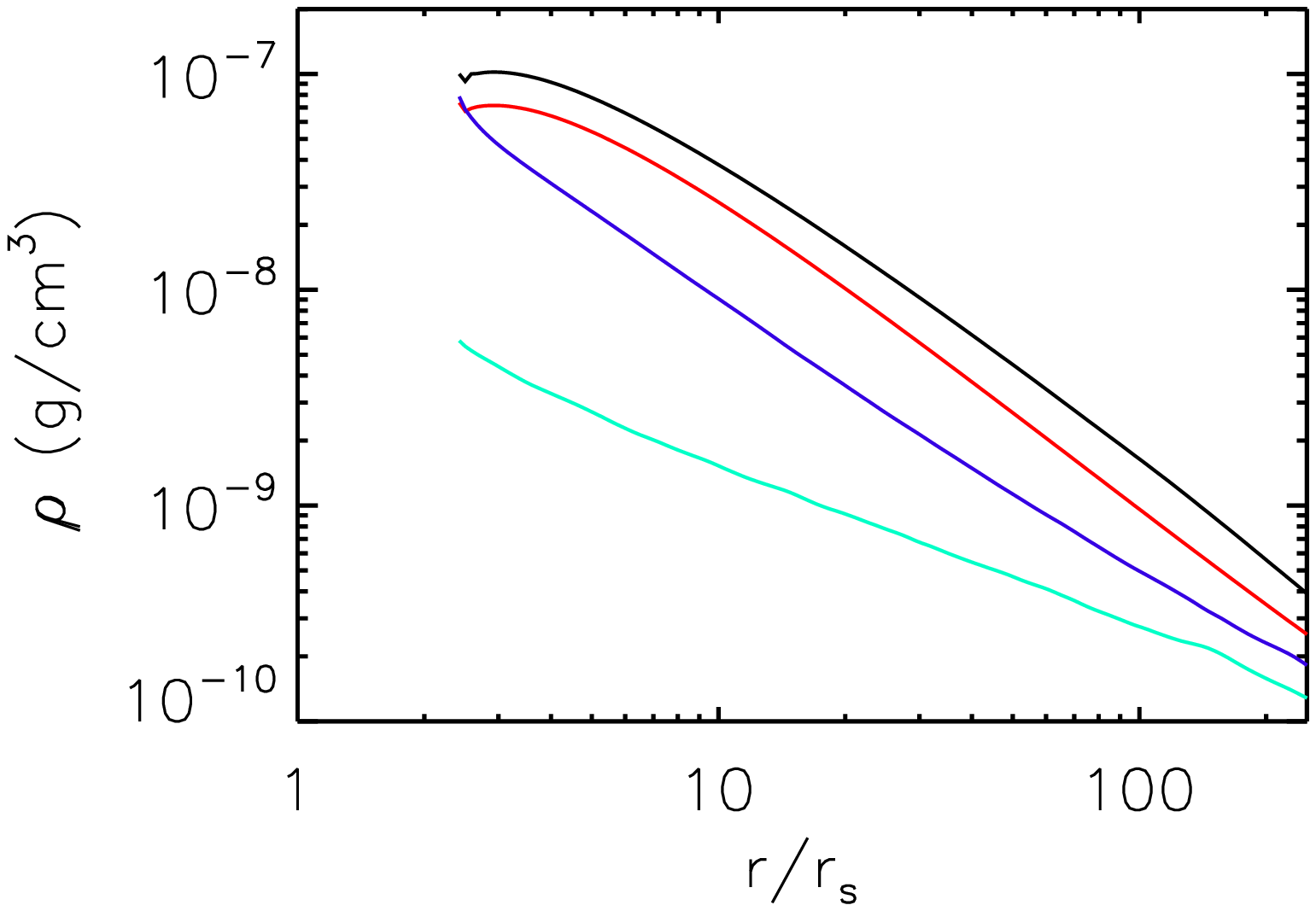}\hspace*{0.7cm}
\includegraphics[scale=0.5]{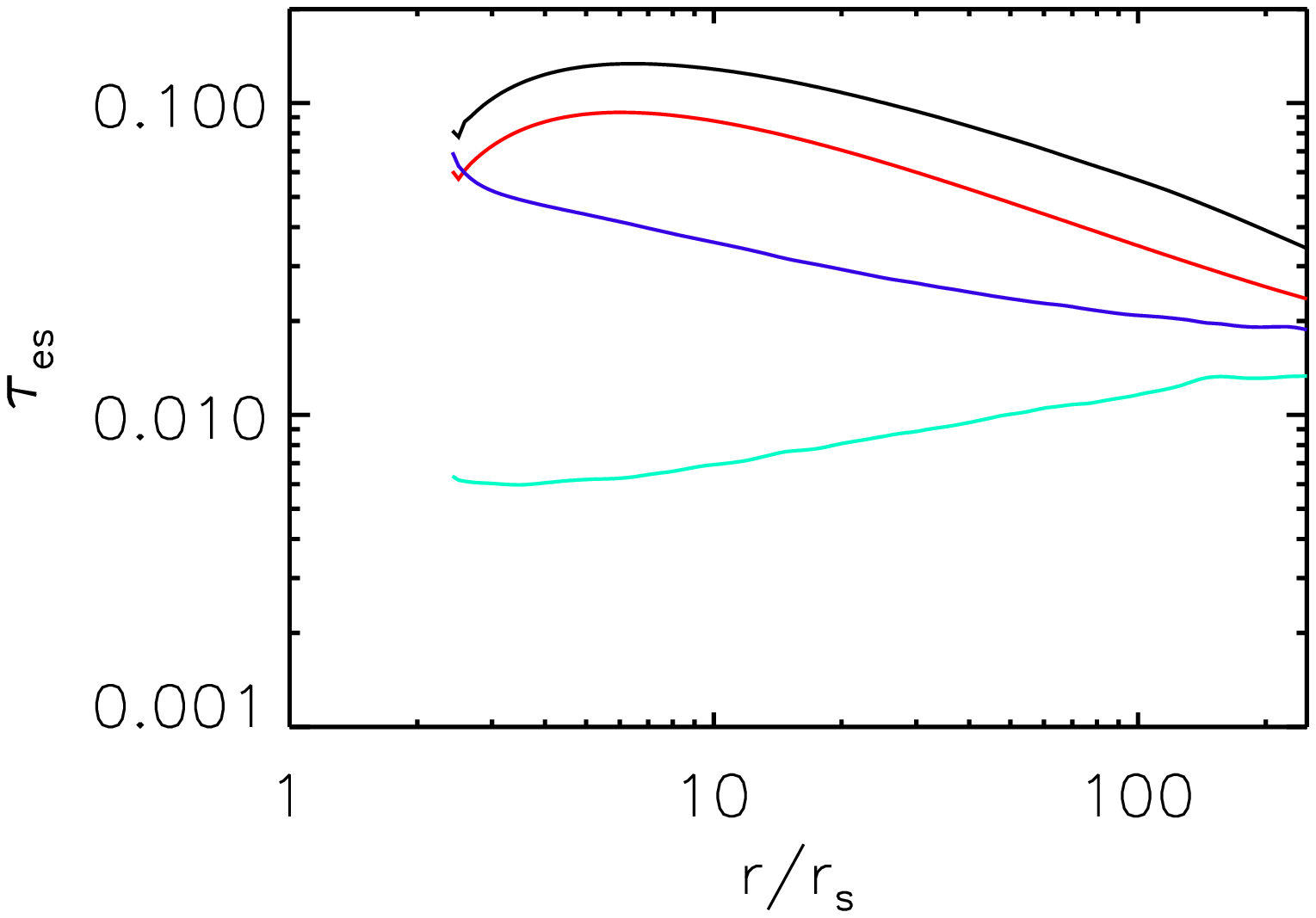}\hspace*{0.0cm} \\
\includegraphics[scale=0.5]{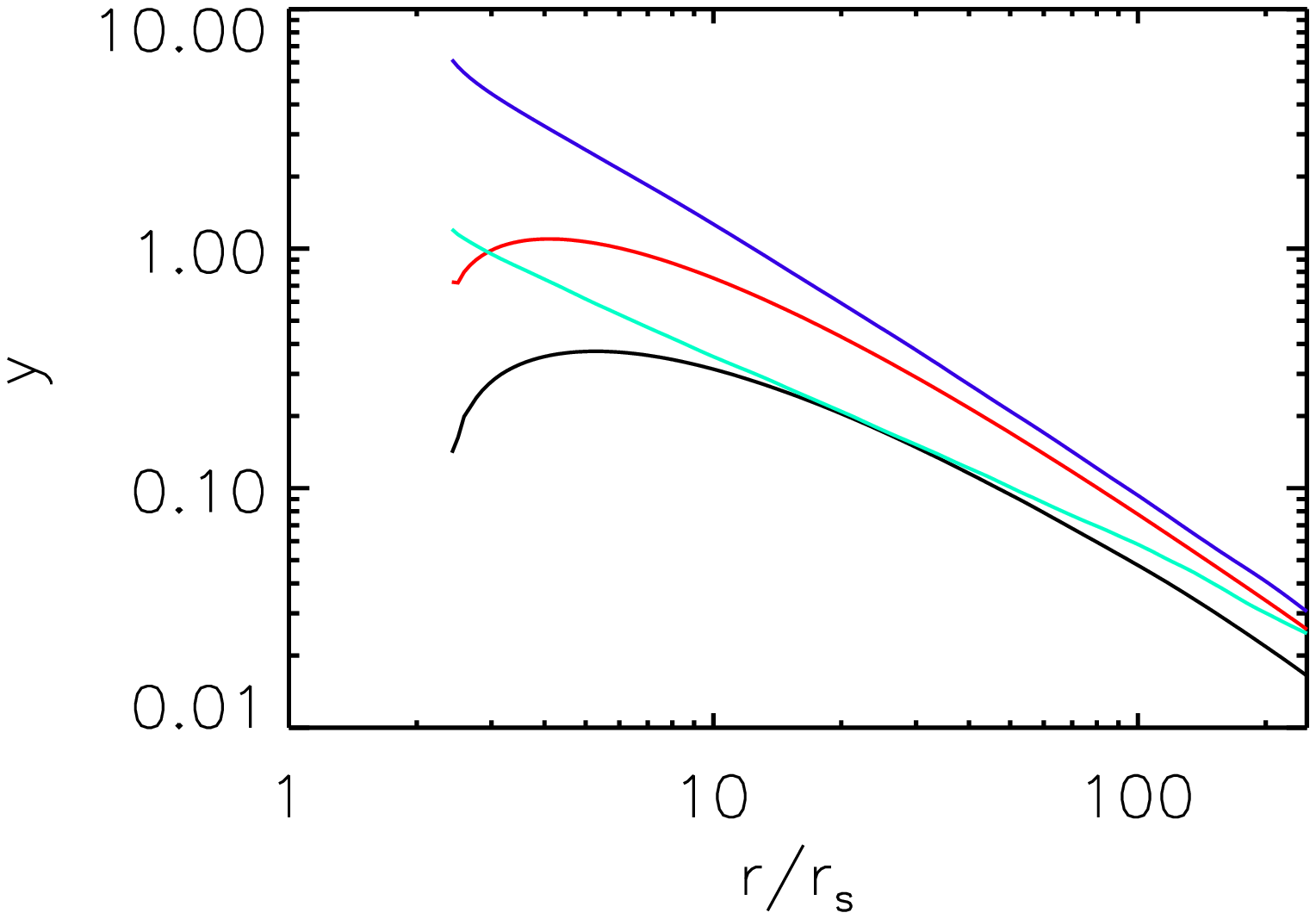}\hspace*{0.7cm}

\hspace*{0cm} \caption{Time-averaged (from $t=3$ to 5 orbital time at initial torus center) variables for models NS1 (black lines, $f_{th}=1$), NS5 (red lines, $f_{th}=0.25$), NS6 (blue lines, $f_{th}=0.05$) and NS7 (green lines, $f_{th}=0.01$). Top left panel: radial profiles of mass inflow (see Equation (\ref{inflowrate})) rate. Top right panel: mass weighted gas temperatures as a function of radius. Middle left panel: radial profiles of volume weighted density. Middle right panel: Compton scattering optical depths as a function of radius (Equation (\ref{tao})). Bottom panel: Compton $y-$ parameters as a function of radius. \label{Fig:NSfth}}
\end{center}
\end{figure*}

\subsection{Effects of $f_{th}$}
In this subsection, we study the effects of $f_{th}$. The initial conditions in models NS1, NS5, NS6 and NS7 are same. The only difference among these four models lies in the value of $f_{th}$ (see Table 1). We show the results in Figure \ref{Fig:NSfth}.

The top left panel plots the mass inflow rate profiles. The initial conditions in these four models are same. Therefore, the mass supply rate at the initial torus location does not differ much. We can see that at 250$r_s$, the mass inflow rates in these models differ by a factor $< 3$. When $f_{th}\geq 0.05$, the strong Compton cooling can reduce wind mass flux. Therefore, in models NS1, NS5 and NS6, the mass inflow rate is almost a constant with radius. In these three models, the mass inflow rate at inner radial boundary is smaller than that at 250 $r_s$ by a factor smaller than $\sim 2$. When $f_{th} < 0.05$, for example in model NS7, the Compton cooling effects are small. In this model, wind is strong and mass inflow rate decreases inwards. The mass inflow rate at inner radial boundary is roughly 1 order of magnitude smaller than that at 250 $r_s$. Correspondingly, in model NS7, the radial density profile is much flatter than those in models NS1, NS5 and NS7 (see middle left panel of Figure \ref{Fig:NSfth}).

With the decrease of $f_{th}$, the gas temperature increases as shown in top right panel of Figure \ref{Fig:NSfth}. The Compton scattering optical depth $\tau_{es}$ decreases with decreasing $f_{th}$. This is because gas density decreases with decreasing $f_{th}$ (see middle left panel of Figure \ref{Fig:NSfth}). Qiao \& Liu (2018) also find that with a same mass accretion rate, $\tau_{es}$ decreases with decreasing $f_{th}$. In models NS1, NS5 and NS6, the densities do not differ much. However, the temperatures increase significantly from models NS1 to NS6. Therefore, the Compton $y-$parameter increases from models NS1 to NS6. Although gas temperature in model NS7 is the highest, the density in this model is much smaller than that in other models. Therefore, the Compton $y-$parameter in model NS7 is not the largest one.

\section{Discussions}
It is well known that a NS has a hard surface. When accretion flow arrives at the surface of a NS, there should be interaction between accretion flow and the surface of the NS. The interaction is complicated. One of the most interesting and difficult topics is that how to decrease the rotational velocity of the accretion disk/flow to the value equal to that of the NS. This topic is studied for both the HAF (Medvedev \& Narayan 2001; Medvedev 2004) and cold thin disk (Gilfanov \& Sunyaev 2014; Popham \& Narayan 1992; Narayan \& Popham 1993; Inogamov \& Sunyaev 1999; Popham \& Sunyaev 2001; Burke et al. 2018). In this paper, we simplify the interaction between the HAF and the surface of a NS. We just assume that the interaction will thermalize the energy carried by the HAF. Also, we assume that a fraction of the thermalized energy will be radiated out as thermal photons. The photons will propagate radially in the accretion flow and cool the accretion flow via Comptonization.

The Maxwell stress is responsible for angular momentum transfer. If the magnetic field is weak and turbulent, angular momentum is transferred by magnetohydrodynamic turbulence induced by magneto-rotational instability (e.g., Balbus \& Hawley 1991). In this paper, we use a viscous stress tensor to transfer the angular momentum. This method should be good enough to mimic angular momentum transfer mechanism by weak turbulent magnetic field. Observations show that generally, magnetic field in NS-LMXBs is very weak ($<10^8$ Gauss; Vaughan et al. 1994; Maccarone \& Coppi 2003; Done et al. 2007). Therefore, our simulations can be applied to most of the NS-LMXBs. There may be some neutron star X-ray binaries with strong large scale magnetic field. If strong large scale magnetic field is present, the accretion flow will be in the magnetically arrested stage (MAD). For MAD, angular momentum is transferred by magnetic braking mechanism (e.g., Tchekhovskoy et al. 2011; McKinney et al. 2012, 2015; Avara et al. 2016; Marshall et al. 2018; Morales et al. 2018). In future, it is very interesting to study the HAF around a NS with strong large scale magnetic field.

In this paper, we neglect the radiative cooling of the accretion flow itself (bremsstrahlung, synchrotron and the Comptonization). For convenience, we define the energy flux of photons radiated out by bremsstrahlung and synchrotron radiation of the accretion flow as $F1$. For the neutron star system, we assume that the energy carried by hot accretion flow advected onto the surface of the neutron star is thermalized as black body emission. We define the energy flux of photons emitted from the surface of the neutron star as $F2$. In the present paper, for both the neutron star and black hole systems, we assume that $F1=0$. For the neutron star system, we take into account the cooling of the accretion flow due to the inverse Compton scattering of $F2$ into the accretion flow. This is the key point in the present numerical simulations. Specifically, we pay attention to the effects of the Compton cooling due to the presence of $F2$ on the dynamics of the accretion flow (e.g., outflow generation and mass accretion rate profiles). For hot accretion flow, most of the viscously dissipated energy is stored in the gas and advected to the central object. Therefore, $F2$ is significantly higher than $F1$. The inverse Compton cooling of $F2$ in the accretion flow dominates that the Compton cooling of $F1$ in the accretion flow. Therefore, the neglect of $F1$ and the corresponding Compton cooling in the present work is reasonable. 

The other assumption in this paper is that ions and electrons have same temperatures. However, for hot accretion flow, two-temperature model is more realistic than one-temperature model (Yuan \& Narayan 2014; Ryan et al. 2017; Sadowski et al. 2017; Bu \& Gan 2018). Esin et al. (1996) did a comparison between one- and two- temperature hot accretion flow models around black holes. It is found that, for a same accretion rate, one-temperature model has a higher radiative efficiency than that of the two-temperature model, that is, at a same accretion rate, one-temperature model is more luminous than two-temperature model. For the neutron star systems, we think that the one-temperature model is a good approximation. This is because the accretion flow around the neutron star is always radiatively efficient due to the existence of the hard surface of the neutron star. Radiatively efficient flow tends to be one-temperature. The detailed comparison of the difference between the one-temperature hot accretion flow and the two-temperature hot accretion flow is unclear, which will be studied in the future work.

We note that the observational results of the scattering optical depth and Compton y-parameter of the accretion flow around neutron stars in some literatures (e.g., Burke et al. 2017, MNRAS, 466, 194) are indeed higher than that of the results in the present work. However, the sources in Burke et al. (2017) have relatively higher luminosities than that of model NS1. Specifically, the sources studied by Burke et al. (2017) have luminosities higher than $1\%$ Eddington luminosity. While in model NS1, the accretion rate is 0.0018 times Eddington accretion rate. If we assume radiative efficiency is 0.1, the luminosity of the accretion flow in model NS1 is 0.0018 times Eddington luminosity. In other calculations in the present numerical simulations, the accretion rate is even lower. Consequently, it is reasonable to have a lower scattering optical depth and Compton y-parameter in the present paper compared with that of Burke et al. (2017).

\section{Summary}
In this paper, we perform two-dimensional hydrodynamic simulations to study the properties of hot accretion flow around a NS. We assume that at the surface of the NS, a fraction of the energy carried by the accretion flow will be thermalized and finally radiated out as the black body emission. The soft photons propagate outward through the accretion flow and cool the accretion flow via Comptonization. We find that the Compton cooling can affect the properties of the HAF around a NS significantly. The effects of Compton cooling become smaller and smaller with decreasing mass accretion rate. When $\dot m>10^{-4}$, Compton cooling can effectively cool the accretion flow and suppress the wind mass flux. Therefore, the mass accretion rate is almost a constant with radius. The density profile is $\rho \propto r^{-1.4}$. When $\dot m < 10^{-4}$, the Compton cooling effects become smaller, wind becomes stronger, the mass inflow rate $\dot M_{\rm in} \propto r^{0.3-0.5}$. Correspondingly, $\rho \propto r^{-1 \sim -0.8}$. When $\dot m < 10^{-6}$, the Compton cooling effects can nearly be neglected. For a same accretion rate, the Compton $y-$parameter of HAF around a NS is lower than that of HAF around a BH. Therefore, theoretically, the HAF around a NS will predict a softer X-ray spectrum compared with that of a BH, which is consistent with observations.

\section*{Acknowledgments}
D. Bu is supported in part by the Natural Science Foundation of China (grants  11773053,
11573051, 11633006 and 11661161012), the Natural Science
Foundation of Shanghai (grant 16ZR1442200), and the Key
Research Program of Frontier Sciences of CAS (No. QYZDJSSW-
SYS008). E. Qiao is supported by the Natural Science Foundation of China (grant 11773037). This work made use of the High Performance Computing Resource in the Core
Facility for Advanced Research Computing at Shanghai Astronomical
Observatory.

\end{document}